\documentclass[11pt,a4paper]{article}
\usepackage{amsfonts}
\usepackage{epsfig}
\usepackage{amsmath}
\usepackage{amssymb}
\usepackage{color}
\usepackage{cite}

\usepackage[dvipdfm,CJKbookmarks,bookmarksopen=true,colorlinks=true,
linkcolor=blue,citecolor=blue,pdfstartview=FitH,pdftitle=title,pdfauthor=lixx]{hyperref}


\begin{document}%
\title{Steady-state traffic flow on a ring road with up- and down- slopes}
\author{ Chun-Xiu Wu$^{1,3}$
\quad Peng Zhang$^{1,3}$\thanks{Corresponding author. E-mail: pzhang@mail.shu.edu.cn.}~ S.C. Wong$^2$~ Keechoo Choi$^4$
\\
\small 1. Shanghai Institute of Applied Mathematics and Mechanics, Shanghai University,\\
\small Shanghai, P.R. China  \\
\small 2. Department of Civil Engineering, The University of Hong Kong, Pokfulam Road,\\
\small Hong Kong SAR, P.R. China\\
\small 3. Shanghai Key Laboratory of Mechanics in
Energy Engineering\\
\small 4 Department of Transportation Engineering, TOD-based Sustainable Urban Transportation Center,\\
 Ajou University, Korea
}
\date{ }%

\maketitle


\noindent \textbf{Abstract}: This paper studies steady-state traffic flow on a ring road with up- and down- slopes using a semi-discrete model. By exploiting the relations between the semi-discrete and the continuum models, a steady-state solution is uniquely determined for a given total number of vehicles on the ring road. The solution is exact and always stable with respect to the first-order continuum model, whereas it is a good approximation with respect to the semi-discrete model provided that the involved equilibrium constant states are linearly stable. In an otherwise case, the instability of one or more equilibria could trigger stop-and-go waves propagating in certain road sections or throughout the ring road. The indicated results are reasonable and thus physically significant for a better understanding of real traffic flow on an inhomogeneous road.
\\


\section{Introduction}
Traffic flow problem with inhomogeneous road conditions has been studied by using the macroscopic continuum model and the microscopic car-following model. In the former formulation, the Riemann problem was concerned mainly for designing a numerical scheme, which could be related to the theory of hyperbolic conservation laws with discontinuous fluxes \cite{Jin:2003,Zhang:2003,Zhang:2005,Zhang:2008,Zhang:2009,Burger:2008a,Burger:2008b,Sun:2011}. In the latter formulation, steady-state flow was analytically or numerically studied by assuming or implying that the solution is piecewise equilibrium constant in the inhomogeneous road sections \cite{Nagai:2006,Tanaka:2006,Hanaura:2007,Ward:2007,Li:2008,He:2009}.

Actually, such a steady-state solution could be described by using the theory of hyperbolic conservation laws with discontinuous fluxes \cite{Zhang:2009,Ward:2007}, which is exact and stable for the first-order continuum LWR model \cite{Lighthill:1955,Richards:1956}. However, it is only an approximate solution to the car-following model or the higher-order model with relaxation if the relaxation time is sufficiently small; otherwise, it is unnecessarily stable due to the underlying metastability in these models. Nevertheless, this feature has not yet been well realized or emphasized in the aforementioned studies of the car-following model.

In the context, the present paper discusses the problem by using a semi-discrete model, which can be viewed as an extension of and thus is more general than the car-following model \cite{Zhang:2012}. On the one hand, we enhance the mathematical analysis of steady-state flow on a ring road with up- and down- slopes, which poses more complexity for the solution. On the other hand, the instability of the solution is emphasized with physical interpretation through analysis of all equilibrium states as well as numerical simulation. Actually, equilibrium traffic flow with an intermediate density value is widely regarded as unstable on a homogeneous ring road, which with oscillations is liable to evolve into stop-and-go waves (e.g., see \cite{Kerner:1994,Zhang:2006,Xu:2007,HMZhang:2003}). Although the discussed ring road is much more complicated, and the simulation does not completely agree with the analytical results due to errors in the analysis, the stead-state solution is indicated to have similar tendency. That is, it is highly stable for a large or small number of vehicles on the ring road; it is unstable for the case that is between. This result is significant for a better understanding of macroscopic properties of traffic flow on an inhomogeneous road. Although not directly applicable, the instabilities could be further associated with those typically indicated in \cite{Kerner:2004,Kerner:2012,Kerner:2013} and those analytically studied in \cite{Seidel:2009,Gasser:2010}.

The remainder of the paper is organized as follows. In Section 2, the semi-discrete model together with its correlation to the continuum model is discussed, and the linear stability condition for an equilibrium solution state is indicated. In Section 3, the wave pattern at a stationary interface is described by using the theory of hyperbolic conservation laws with discontinuous fluxes, which helps determine the two adjacent equilibrium constant states. Accordingly, the wave types of the steady-state solution on a ring road with up- and down- slopes are analytically indicated, which depend on the total number of vehicles on the ring road. In Section 4, initial distributions with certain total number of vehicles are shown to converge to or diverge from the corresponding steady-state solutions through numerical simulation, which agrees with the analytical results in general. We conclude the paper by Section 5.

\section{The semi-discrete model}
In the semi-discrete model \cite{Zhang:2012}, a moving \textquotedblleft particle" in traffic flow could be numbered by $m$ with $x_m(t)$ being its position, and
  \begin{eqnarray}\label{1}
     \frac{dx_m(t)}{dt}=u_m(t),\;m=0,\pm1,\pm2,\cdots
   \end{eqnarray}
being its speed, and the acceleration was defined through
  \begin{eqnarray}\label{2}
    \frac{d}{dt}[u_m(t)+p(s_m(t))]=\frac{1}{\tau}[u_e(s_m(t))-u_m(t)].
  \end{eqnarray}
Here,
\begin{eqnarray}\label{3}
s_m(t)=[x_{m+1}(t)-x_m(t)]/\Delta M,\;\;\text{and}\;\rho_m(t)\equiv(s_m(t))^{-1},
\end{eqnarray}
are the specific volume and the density represented by the particle $m$, $\Delta M$ is the mass between the particles $m$ and $m+1$, $u_e(s)$ and $p(s)$ are the equilibrium velocity-density relationship and the traffic pressure satisfying $u'_e(s)>0$, and $p'(s)\leq 0$. For $\Delta M=1$, the semi-discrete model of (\ref{1}) and (\ref{2}) reduces to a car-following model, in which case each particle can be viewed as a vehicle in that there is just one vehicle between the heads of two adjacent vehicles. The resultant car-following model is essentially the same as that in \cite{Aw:2002,Greenberg:2001,Greenberg:2004}.

\subsection{The correlation to the continuum model in Lagrange coordinates}
Assume that there is no overtake between particles for division with a sufficiently small increment $\Delta M$, and let $M$ denote the total mass not passing through the particle $m$. This implies that $M$ referring to the particle $m$ is independent of time $t$. Therefore, the particle $m$ can be identified by $M$ and an associated variable $A_m(t)$ (e.g., the position $x_m(t)$ and speed $v_m(t)$) can be re-denoted by $A(M,t)$. Let $\Delta M\rightarrow0$, then the flow can be viewed as a continuum, in which $M$ and $t$ constitute the Lagrange coordinate system to describe the variable $A(M,t)$. In this case, Eq. (\ref{3}) gives
   \begin{eqnarray}\label{4}
   s(M,t)=x_M(M,t),\;\;\text{and}~\rho(M,t)=(s(M,t))^{-1}.
   \end{eqnarray}
Furthermore, we reduce Eq. (\ref{1}) from $dx_{m+1}(t)/dt=u_{m+1}(t)$, and divide the resultant equation by $\Delta M$. Then, for $\Delta M\rightarrow0$, we have the following partially differential equation:
   \begin{eqnarray}\label{5}
   s_t(M,t)-u_M(M,t)=0.
   \end{eqnarray}
It is obvious that, for $\Delta M\rightarrow0$, Eq. (\ref{2}) yields
    \begin{eqnarray}\label{6}
   [u(M,t)+p(s(M,t))]_t=\frac{u_e(s(M,t))-u(M,t)}{\tau}.
    \end{eqnarray}
The discussion implies the consistency between the semi-discrete model (\ref{1})-(\ref{2}) and the continuum model (\ref{5})-(\ref{6}), namely, the former system converges to the latter system for $\Delta M\rightarrow 0$. See \cite{Zhang:2012} for more detailed discussion of the Lagrange coordinates.

\subsection{The correlation to the continuum model in Euler coordinates}
Let now $A(M,t)$ be denoted by the Euler coordinates $(x,t)$ with $A(M,t)=A(x,t)$, where $x=x(M,t)$ is the position. We have
   \begin{eqnarray}\label{7}
   A_t(x,t)=A_x(x,t) x_t(M,t)+A_t(x,t),\;A_M(x,t)=A_x(x,t) x_M(M,t),
   \end{eqnarray}
where $x_t(M,t)=u(M,t)=u(x,t)$, and $x_M(M,t)=s(M,t)=s(x,t)=1/\rho(x,t)$, according to Eqs. (\ref{1}) and (\ref{4}). Replacing the partial derivatives of $s(M,t)$ and $u(M,t)$ in Eqs. (\ref{5})-(\ref{6}) through Eq. (\ref{7}), we have
   \begin{eqnarray}\label{8}
    \rho_t+(\rho u)_x=0,\\ \label{9}
    [\rho (u+P(\rho))]_t+[\rho u (u+P(\rho))]_x=\frac{\rho U_e(\rho)-\rho u}{\tau},
   \end{eqnarray}
where $P(\rho)=p(s)$, $U_e(\rho)=u_e(s)$, and $\rho=1/s$. The system of (\ref{8})-(\ref{9}) turns out to be the so called \textquotedblleft anisotropic" higher-order traffic flow models in \cite{Aw:2000,Rascle:2002,HMZhang:2002,HMZhang:2009}. This again establishes the correlation between the microscopic semi-discrete model and the macroscopic continuum model.

The consistence between the aforementioned three systems implies that the solution to the semi-discrete model should converge to the solution to the other two systems for $\Delta M\rightarrow 0$.

\subsection{Stability of equilibrium solution}
 An equilibrium constant state $(\rho_0,U_e(\rho_0))$ or $(s_0,u_e(s_0))$ is linearly stable with respect to the higher-order model (\ref{8})-(\ref{9}) or (\ref{5})-(\ref{6}), if
    \begin{eqnarray}\label{10}
    U'_e(\rho_0)+P'(\rho_0)\geq 0,\;\;\text{or}\;\;u'_e(s_0)+p'(s_0)\leq 0.
    \end{eqnarray}
This corresponds to an equilibrium state
    \begin{eqnarray}\label{11}
   x^0_{m+1}(t)-x^0_m(t)=s_0\Delta M,\;\text{or}\;\; x^0_m(t)=u_e(s_0)t+m\Delta Ms_0,
    \end{eqnarray}
in the semi-discrete model (\ref{1})-(\ref{2}), which is linearly stable if
    \begin{eqnarray}\label{12}
 u'_e(s_0)+p'(s_0)\leq \frac{\Delta M}{2\tau}.
    \end{eqnarray}
As $\Delta M\rightarrow 0$, Eq. (\ref{12}) reduces to Eq. (\ref{10}), which also indicates the consistency between the continuum and the semi-discrete models. See \cite{Zhang:2012} for detailed discussion.

\section{The steady-state solution over inhomogeneous road sections}
For discussion of the steady-state solution, the system (\ref{8})-(\ref{9}) under the Euler coordinates is convenient to deal with the interface between two inhomogeneous road sections. For sufficient small $\tau$, Eq. (\ref{9}) is approximated by $u=U_e(\rho)$, which together with Eq. (\ref{8}) leads to the following LWR model \cite{Lighthill:1955,Richards:1956}:
   \begin{eqnarray}\label{13}
  \rho_t+(Q_e(\rho))_x=0,
   \end{eqnarray}
where $Q_e(\rho)=\rho U_e(\rho)$ is the flow-density relationship or fundamental diagram.

\subsection{General discussion of the solution}
In general, the steady-state solution $(\rho(x,t),u(x,t))=(\rho(x),u(x))$ of (\ref{13}) on an inhomogeneous road is piecewise constant satisfying the equilibrium condition $u(x)=U_e(\rho(x))$. At each dividing point or interface, the flow $q\equiv\rho u=Q_e(\rho)$ is continuous and thus is constant all over the road. However, the density $\rho(x)$ and the velocity $u(x)$ are usually discontinuous at the dividing point, which can be either a contact between two inhomogeneous road sections or a stationary shock in a homogeneous road section.

The constancy of the flow at the dividing point is attributed to the mass conservation or the Rankine-Hugoniot condition (see \cite{Zhang:2003,Zhang:2005,Zhang:2008,Zhang:2009} and the references therein). Let $\rho=\rho_{\mp}$ and $Q_e^{\mp}$ denote the densities and fundamental diagrams on the left and right hand sides of the interface. Then, the mass conservation is simply that
    \begin{eqnarray}\label{14}
       Q_e^-(\rho_-)=Q_e^+(\rho_+).
    \end{eqnarray}
For the dividing point which is stationary, a \textquotedblleft wave entropy" condition can be applied at the interface to describe valid wave breaking on the both sides \cite{Zhang:2003,Zhang:2005,Zhang:2008,Zhang:2009}. That is, the characteristic speeds on the two sides take the same sign with
 \begin{eqnarray}\label{15}
(Q^-_e(\rho_-))'(Q_e^+(\rho_+))'>0.
 \end{eqnarray}
Otherwise, we have
  \begin{eqnarray}\label{16}
  (Q^+_e(\rho_+))'\geq0,\; \text{if}\;(Q^-_e(\rho_-))'=0;
   \end{eqnarray}
and
   \begin{eqnarray}\label{17}
    (Q^-_e(\rho_-))'\leq 0,\; \text{if}\;(Q^+_e(\rho_+))'=0.
  \end{eqnarray}
See also \cite{Burger:2008a,Burger:2008b} for a similar entropy condition.

We assume that the critical densities are $\rho_-^*$ and $\rho_+^*$ with $Q_e(\rho_-^*)$ and $Q_e(\rho_+^*)$ being the maximal flows or capacities on the left and the right hand sides, i.e.,
    \begin{eqnarray}\label{18}
   (Q^{\mp}_e(\rho_\mp^*))'=0;\; (\rho-\rho_\mp^*)(Q^{\mp}_e(\rho))'<0,\;\text{for}\;\rho\neq\rho_\mp^*.
    \end{eqnarray}
For the discussed steady-state solution $\rho=\rho(x)$, we obviously have
    \begin{eqnarray}\label{19}
    Q_e(\rho(x))\leq \min(Q^-_e(\rho_-^*),Q^+_e(\rho_+^*)),
    \end{eqnarray}
and the term on the right hand side is defined as the capacity with respect to the two divided sections. Thus, we have four wave patterns at the interface (Fig. 1). The normal and the congested wave patterns (Figs. 1(a) and (d)) are associated with Eq. (\ref{15}). The upstream or downstream capacitated wave patterns (Fig. 1(b) or (c)) is associated with Eq. (\ref{16}) or (\ref{17}), which occurs only if $Q^-_e(\rho_-^*)\leq Q^+_e(\rho_+^*)$ or vice verse. We call the interface a bottleneck if $Q^-_e(\rho_-^*)>Q^+_e(\rho_+^*)$, which triggers a queue upstream from the interface when the traffic is downstream capacitated.\\

\begin{center}
\scriptsize(a)\epsfig{figure=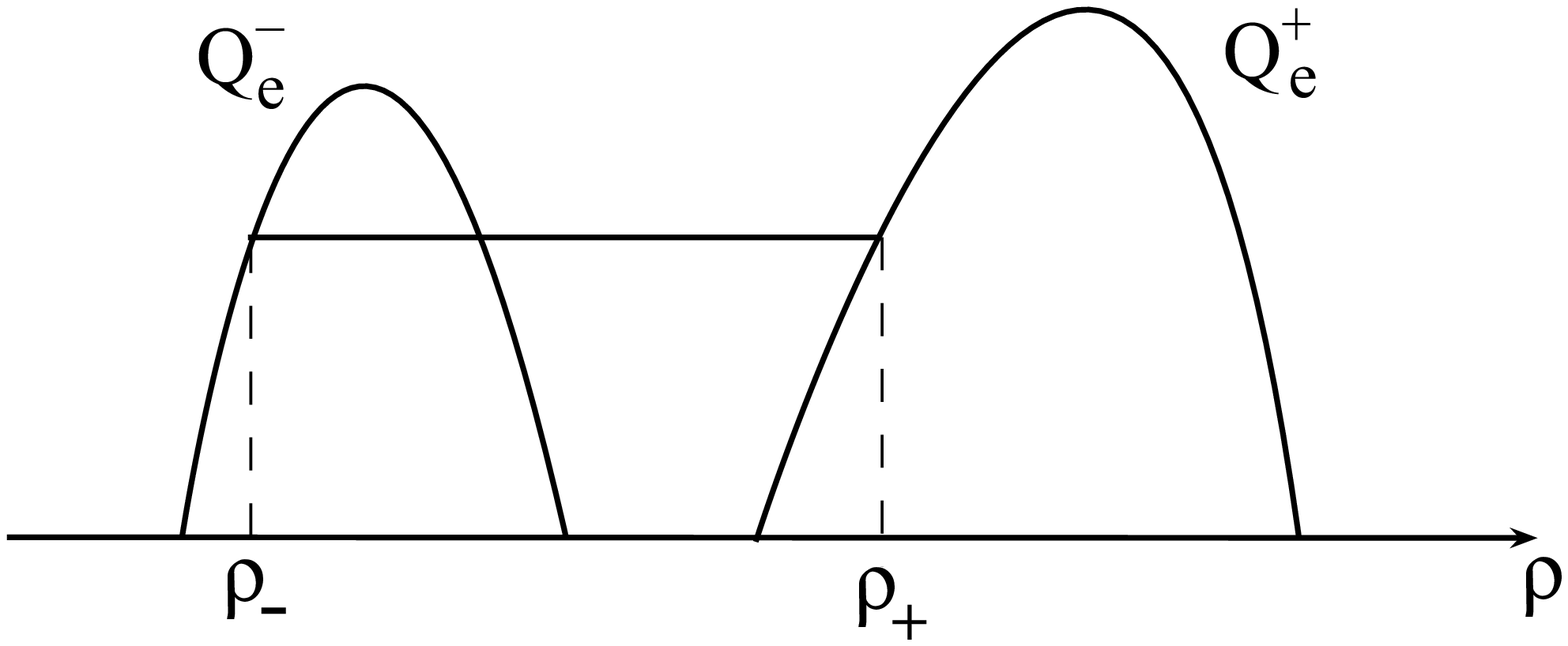,width=2.8 in}  \hspace{0.2 cm}
\scriptsize(b)\epsfig{figure=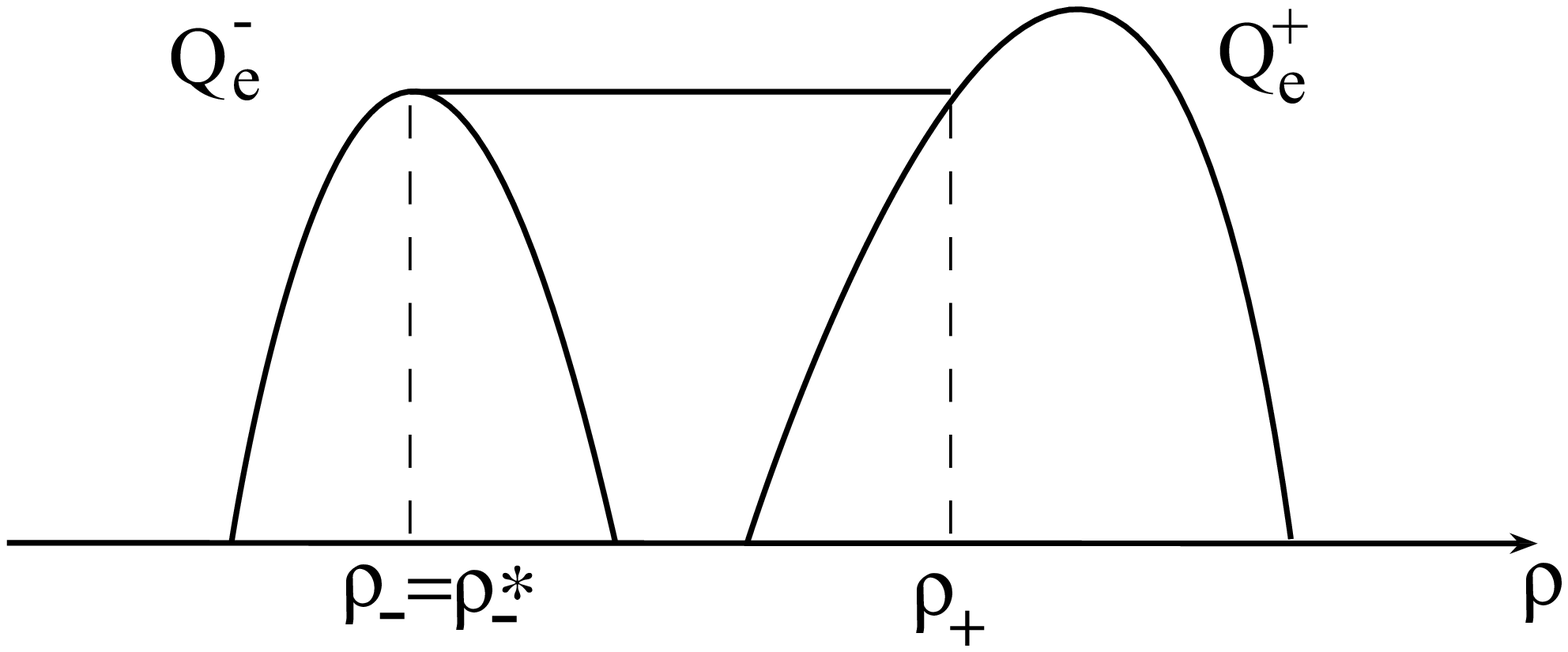,width=2.8 in}  \hspace{0.2 cm}\\
\scriptsize(c)\epsfig{figure=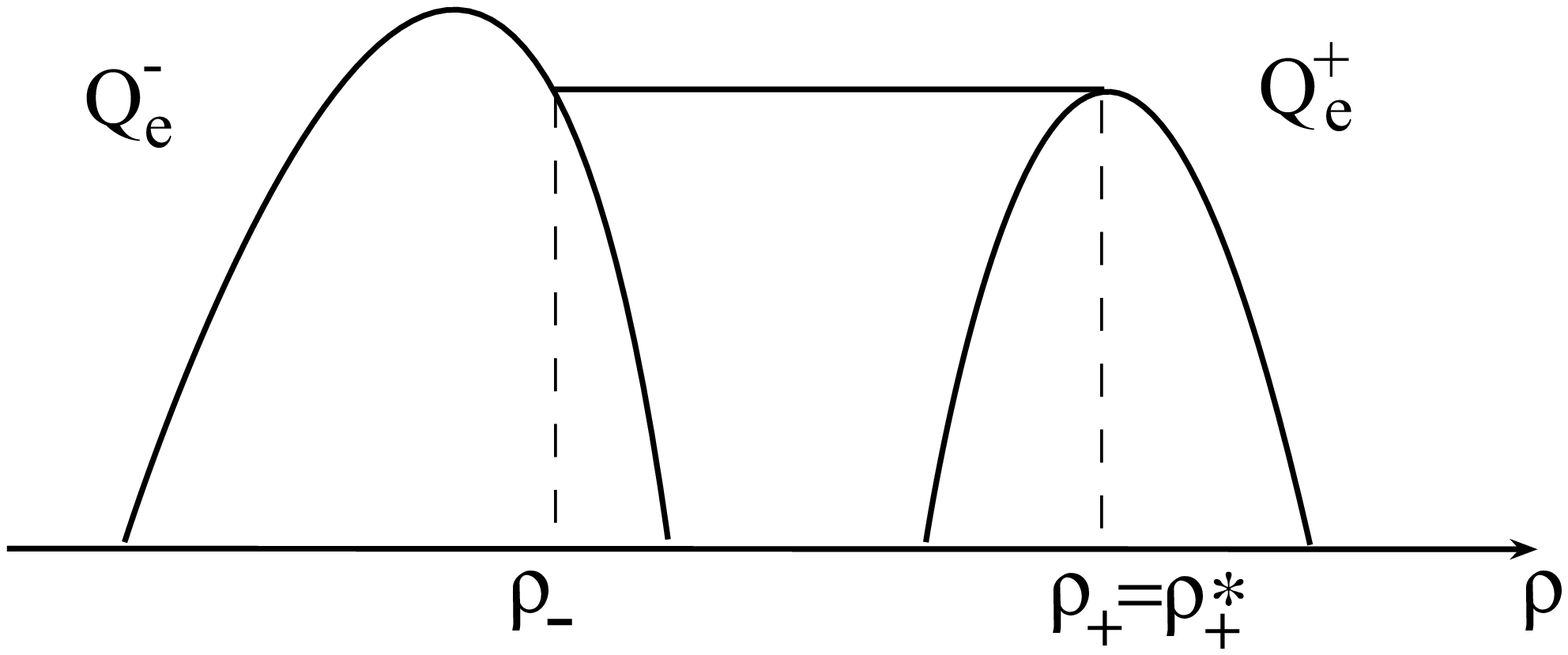,width=2.8 in}  \hspace{0.2 cm}
\scriptsize(d)\epsfig{figure=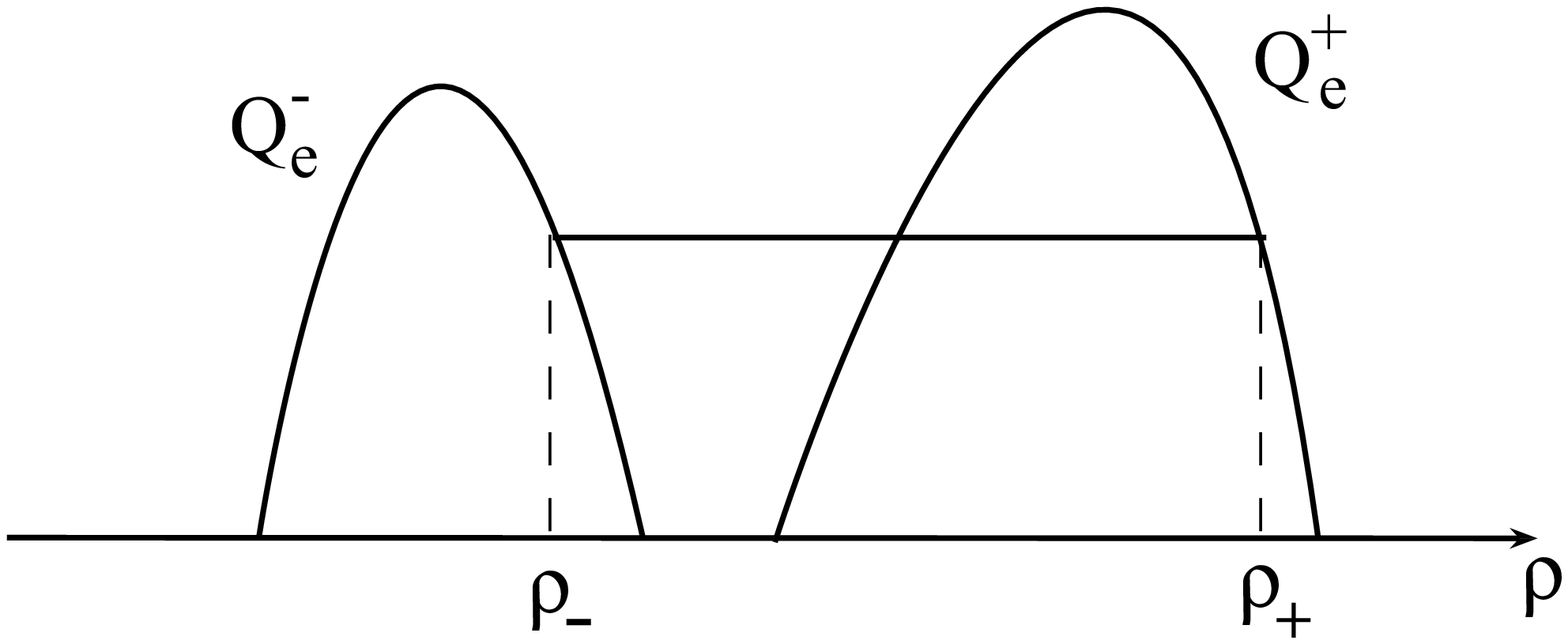,width=2.8 in}  \hspace{0.2 cm}\\
 \vspace{0.3cm}
 {\footnotesize Fig. 1 Four wave patterns at a joint between two inhomogeneous road sections, which obeys the conservation of (\ref{11}). (a) and (d): normal and congested wave patterns associated with Eq. (\ref{15}); (b) and (c): upstream and downstream capacitated wave pattern associated with Eqs. (\ref{16}) and (\ref{17}), respectively.}
\end{center}

\subsection{The solution on a ring road with up- and down- slopes}
Steady-state flow on an inhomogeneous highway road is characterized by the wave pattern at each of the interfaces. For a road with up- and down- slopes, we adopt the following velocity-specific volume  relationship:
\begin{eqnarray}\label{20}
  u_e(s)=\frac{u_f(\beta)[tanh(s/l-x_c(\beta)/l)+tanh(x_c(\beta)/l-1)]}{1+tanh(x_c(\beta)/l-1)},
 \end{eqnarray}
where $\beta$ is the slope, $l$ the vehicle length, and $u_f(\beta)$ the free flow velocity. We can easily examine that $u_e(l)=0$, $u_e(+\infty)=u_f(\beta)$, and $u'_e(s)>0$, for $s\geq l$. By the relations: $U_e(\rho)=u_e(s)$, $\rho=s^{-1}$, we correspondingly have $U'_e(\rho)<0$, for $\rho\in[0,\rho_{jam}]$, and $U_e(\rho_{jam})=0$, $U_e(0)=u_f(\beta)$, where $\rho_{jam}=1/l$ is the jamming density.
By scaling, the free flow velocity $u_f(\beta)$ and the safe interval $x_c(\beta)$ are determined by the following piecewise functions:
\begin{eqnarray}\label{21}
  \frac{u_f(\beta)}{u_f(0)}=\left\{ \begin{array} {cc}
  -100\beta^2-5\beta+1, & -0.10\leq \beta< 0, \\
  1, & 0\leq \beta< 0.02,  \\
  -150\beta^2+3\beta+1, & 0.02\leq \beta\leq 0.08, \\
  0.28,&0.08< \beta\leq 0.10,  \\
  \end{array}\right.\;\;u_f(0)=30~m/s,
 \end{eqnarray}
and
   \begin{eqnarray}\label{22}
  \frac{x_c(\beta)}{l}=\left\{ \begin{array} {cc}
  300\beta^2-12\beta+3, & -0.10\leq \beta< 0,\\
  80\beta^2+15\beta+3, & 0\leq \beta\leq 0.10,\\
  \end{array}\right.\;\;l=4.5~m,
\end{eqnarray}
The formula (\ref{20}) is extended from that in \cite{Greenberg:2004}, and Eqs. (\ref{21})-(\ref{22}) are based on the experimental study in \cite{Zhou:2004}. The formula (\ref{20}) can also be viewed as a modification of that in \cite{Li:2008}, which was extended from that in \cite{Bando:1995}. We note that the properties of (\ref{18}) can be verified for $|\beta|\leq 0.1$.\\

\begin{center}
\scriptsize(a)\epsfig{figure=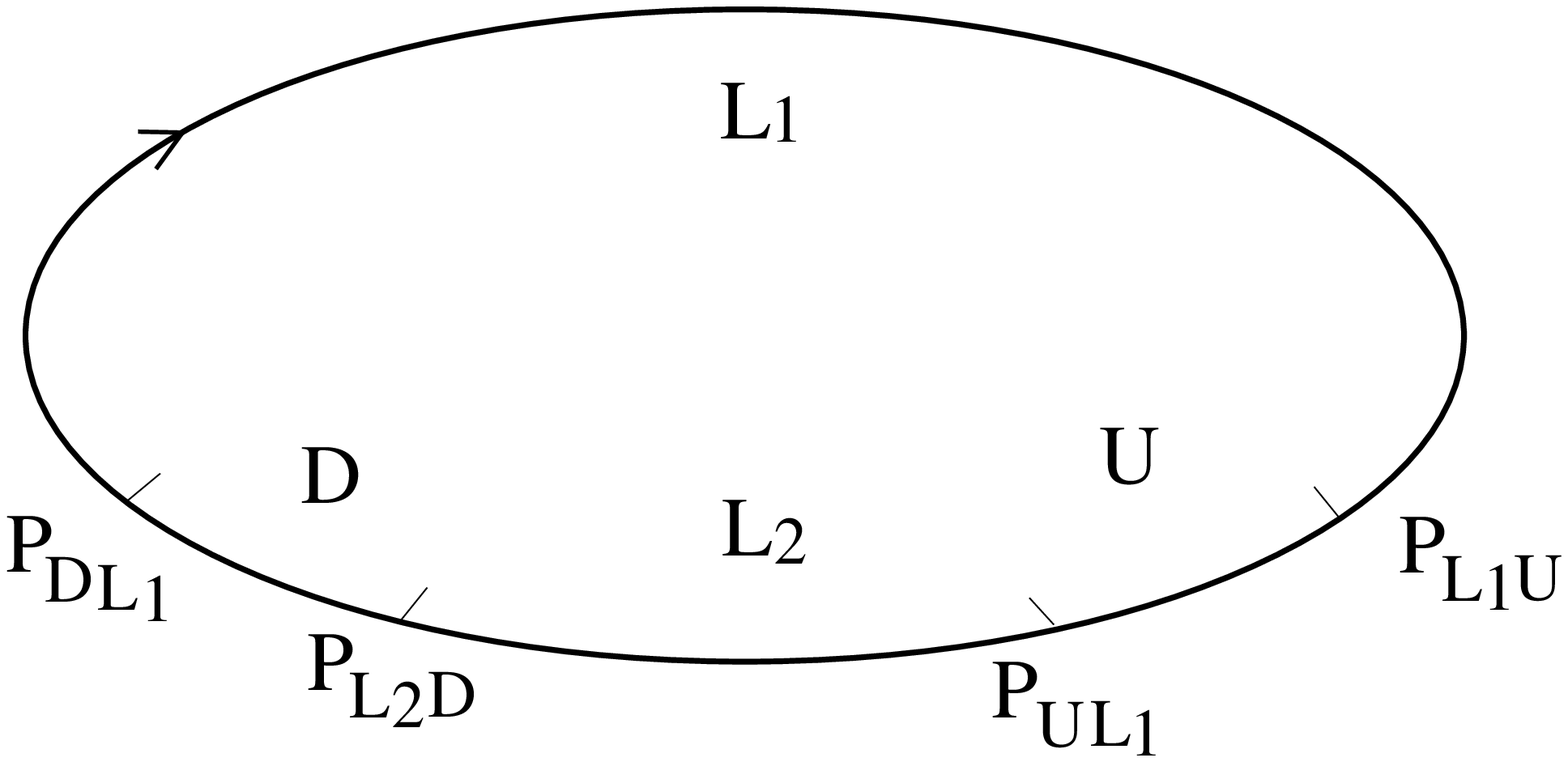,width=2.8 in}  \hspace{0.0 cm}
\scriptsize(b)\epsfig{figure=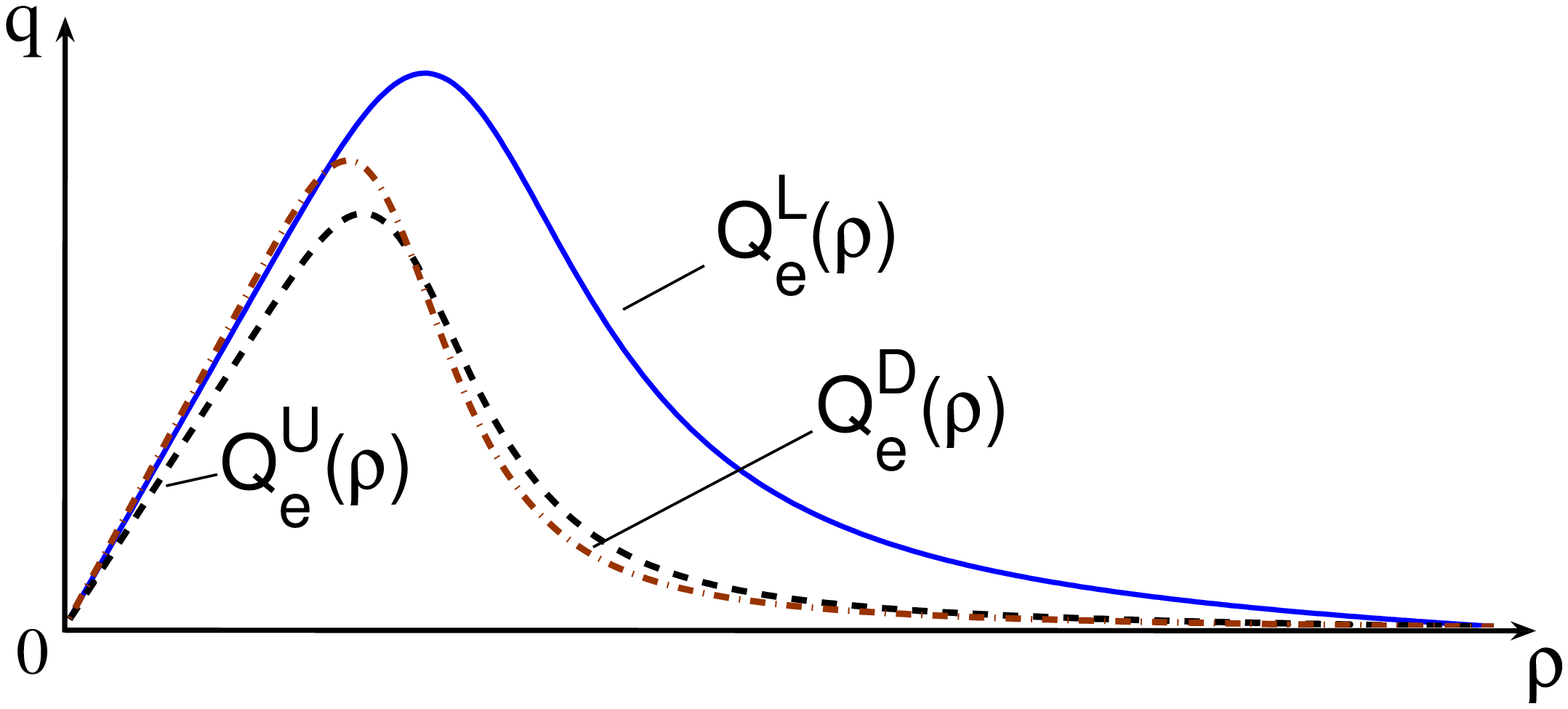,width=2.8 in}  \hspace{0.0 cm}\\
\vspace{0.3cm}
 {\footnotesize Fig. 2 (a) The clockwise ring road $R=L_1\cup U\cup L_2\cup D$, with the lengths $|R|=6750m$, $|L_1|=4050m$, $|U|=675m$, and $|D|=675m$; (b) the fundamental diagrams $q=Q_e^L(\rho)$, $q=Q_e^U(\rho)$, and $q=Q_e^D(\rho)$, for the level road sections $L_1\cup L_2$, the up-slope section $U$, and the down-slope section $D$, which are defined through Eqs. (\ref{20})-(\ref{22}).}
\end{center}

We now consider a clockwise ring road $R$ which is composed of four sections: (i) the level road $L_1$ with $\beta=0$; the up-slope $U$ with $\beta=0.04$; the level road $L_2$ with $\beta=0$; and the down-slope $D$ with $\beta=-0.04$. The joints between two adjacent sections are denoted by $P_{L_1U}$, $P_{UL_2}$, $P_{L_2D}$, and $P_{DL_1}$. The ring road is shown in Fig. 2(a), and the fundamental diagrams $q=Q_e(\rho)$ for all sections are shown in Fig. 2(b). We see that, by Fig. 2(b),
   \begin{eqnarray}\label{23}
   Q_s\leq\min(Q^L_e(\rho_L^*),Q^U_e(\rho_U^*),Q^D_e(\rho_D^*))=Q^U_e(\rho_U^*)\equiv Q_c,
   \end{eqnarray}
where $Q_s=Q_e(\rho(x))$ is the flow corresponding to the steady-state solution $\rho(x)$, and $Q_c$ is defined as the capacity of the ring road. Eq. (\ref{23}) is the extension of Eq. (\ref{19}). Given $Q_s$, we have two solution values of $\rho(x)$ through Eqs. (\ref{20})-(\ref{22}). However, $\rho(x)$ can be uniquely determined according to the wave pattern at each dividing point. Actually, $Q_s$ together with the wave patterns can be uniquely determined by the total number $N$ of vehicles on the ring road, which is indicated in the following discussion by referring to Figs. 2(a) and (b).

For $N$ increasing from $N=0$, $Q_s$ increases until it reaches the capacity with $Q_s=Q_c$, when the total number of vehicles
\begin{eqnarray}\label{24}
     N_{t_1}=\sum_\chi|\chi|\rho_\chi,\;\chi=L_1,U,L_2,D.
   \end{eqnarray}
Here, $|\chi|$ and $\rho_\chi$ represent the lengths of and the densities in four sections divided by $P_{L_1U}$, $P_{UL_2}$, $P_{L_2D}$, and $P_{DL_1}$. In this case, we have $\rho_\chi<\rho_\chi^*$, except for $\rho_U=\rho_U^*$.

For $N<N_{t_1}$, we have $Q_s<Q_c$, and $\rho_\chi<\rho_\chi^*$, for all $\chi$, thus the interfaces at all dividing points represent normal wave patterns, which are associated with Eqs. (\ref{11})-(\ref{12}). Accordingly, the constant flow $Q_s$ together with the densities $\rho_\chi$ in the four sections is uniquely determined by
   \begin{eqnarray}\label{25}
    \sum_\chi|\chi|\rho_{\chi}=N,\;Q_e(\rho_\chi)=Q_s,\;\chi=L_1,U,L_2,D.
   \end{eqnarray}

For $N> N_{t_1}$, we have the same flow rate $Q_s=Q_c$, and other density values as those in the case of $N= N_{t_1}$, except for a \textquotedblleft blow-up" upstream from $P_{L_1U}$, which is due to the capacity drop $Q_e(\rho^*_U)=Q_c<Q_e(\rho^*_L)$ (see Eq. (\ref{23}) and Fig. 2(b)) at this interface. According to the discussion in Section 3.1, this suggests that $P_{L_1U}$ represent the one and only bottleneck on the whole road with respect to the discussed solution. In this case, $L_1$ is divided into two intervals $L_{11}$ and $L_{12}$, and their lengths together with the dividing point are determined by $N$ through the following equation:
    \begin{eqnarray}\label{26}
    N=\sum_\chi |\chi|\rho_{\chi},\;\chi=L_{11},L_{12},U,L_2,D.
    \end{eqnarray}
In this case, the traffic is downstream capacitated at $P_{L_1U}$, and $P_{L_{11}L_{12}}$ represents the position $P_s$ of a stationary shock; the threshold value of $N$ is computed by
   \begin{eqnarray}\label{27}
     N_{t_2}=\sum_\chi |\chi|\rho_{\chi},\;\chi=L_1,U,L_2,D,
   \end{eqnarray}
when $P_s$ coincides with $P_{DL_1}$.

As $N$ continues to increase, the position $P_s$ of the stationary shock moves upstream until it reaches the joint $P_{UL_2}$, which is between the up-slope and the level road 2. In this process, a joint becomes to represent a congested wave pattern if its position is downstream from $P_s$. We have the threshold value
    \begin{eqnarray}\label{28}
     N_{t_3}=\sum_\chi |\chi|\rho_{\chi},\;\chi=L_1,U,L_2,D,
   \end{eqnarray}
when $P_s$ coincides with $P_{L_2D}$, and the threshold value
    \begin{eqnarray}\label{29}
     N_{t_4}=\sum_\chi |\chi|\rho_{\chi},\;\chi=L_1,U,L_2,D,
   \end{eqnarray}
when $P_s$ coincides with $P_{UL_2}$. The solution $\rho(x)$ together with the position $P_s$ of the stationary shock for $N$ between $N_{t_i}$ and $N_{t_{i+1}}$ ($i=2,3$) can be determined similarly to Eq. (\ref{26}).

For $N>N_{t_4}$, the constant flow $Q_s$ cannot reach the capacity $Q_c$ of the ring road. As $N$ increases, $Q_s$ begins to decrease until $Q_s=0$, when the whole road becomes completely blocked. In this development, $Q_s$ together with $\rho_{\chi}$ in the four sections is determined also by Eq. (\ref{25}), except that each joint between two adjacent road sections represents a congested wave pattern.

The function $Q_s=Q_s(N)$ is shown in Fig. 3(a). Correspondingly, the characteristics of the wave patterns at all joints are shown in Fig. 3(b).\\
\begin{center}
\scriptsize(a)\epsfig{figure=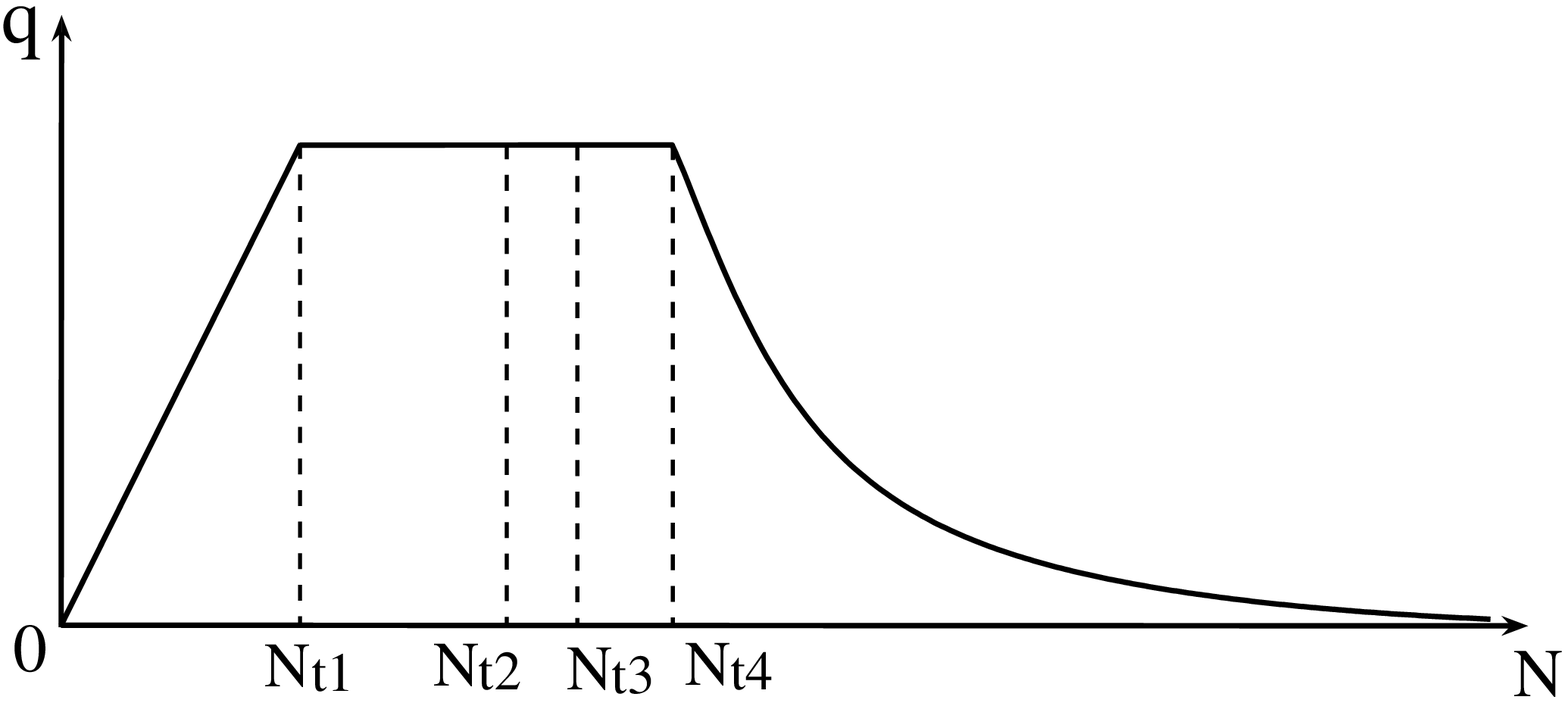,width=2.8 in}  \hspace{0.0 cm}
\scriptsize(b)\epsfig{figure=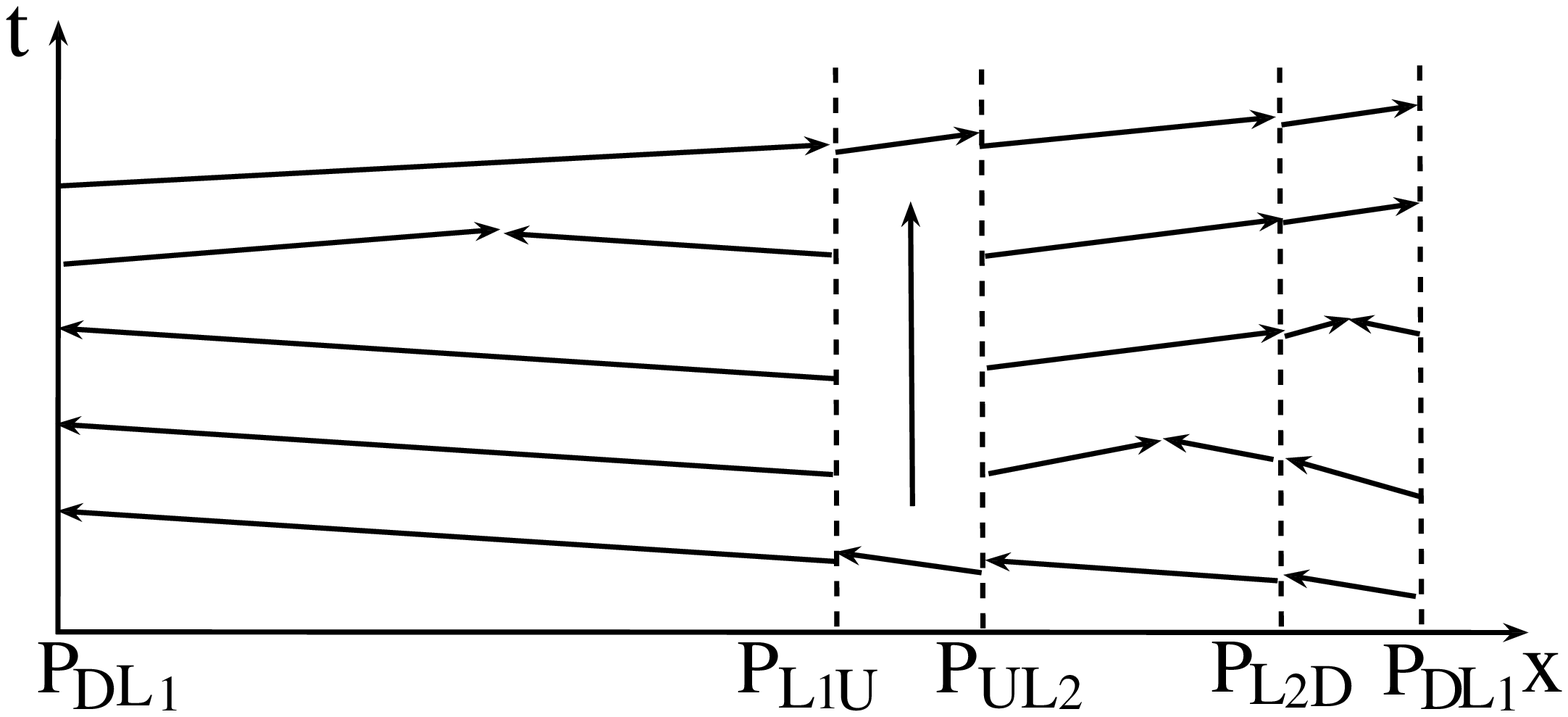,width=2.8 in}  \hspace{0.0 cm}\\
\vspace{0.3cm}
 {\footnotesize
 Fig. 3 (a) The function $Q_s=Q_s(N)$; (b) resultant wave patterns at all joints for $N<N_{t_1}$, $N_{t_1}<N<N_{t_2}$, $N_{t_2}<N<N_{t_3}$, $N_{t_3}<N<N_{t_4}$, and $N>N_{t_4}$, from the top to the bottom, respectively, where the left, up and right arrows represent negative, zero and positive characteristic speeds, respectively, and a position intersected by a left arrow and a right arrow represents a stationary shock.
 }
\end{center}

\section{Numerical simulation}
The semi-discrete model of (\ref{1})-(\ref{2}) is used for numerical simulation. We casually set the following initial distribution:
    \begin{eqnarray}\label{30}
     s_0=\frac{|R|}{N},\;x_m(0)=ms_0,\;v_m(0)=u_e(s_0),\;m=1,\cdots,N,
    \end{eqnarray}
and define $s_N(t)=(x_1(t)+|R|-x_N(t))/\Delta M$ in Eq. (\ref{2}), where the length of the ring road $|R|=6750m$. In the simulation, the position that is $2700$ meters downstream from the joint $P_{L_1U}$ is taken as the origin. Moreover, the length of a road section is scaled by $l=4.5m$ and the density is scaled by $\rho_{jam}=1/l$. As a consequence, the coordinates of the joints $P_{L_1U}$, $P_{UL_2}$, $P_{L_2D}$, and $P_{DL_1}$ are $600$, $750$, $1050$, and $1200$, respectively.

Since $u_e(s_0)$ in Eq. (\ref{30}) depends on the slope $\beta$, the vehicular velocities in different road sections are not equal. As a consequence, the headway cannot keep constant in the evolution. Then, given the total number $N$ of vehicles on the ring road, the simulation is designed to test whether the initial distribution of (\ref{30}) converges to the corresponding steady-state solution. Since the simulation involves a stiff relaxation with small $\tau$, a semi-implicit scheme for time discretization of (\ref{1})-(\ref{2}) is adopted,
     \begin{eqnarray}\label{31}
     \frac{x_m^{(n+1)}-x_m^{(n)}}{\Delta t}=u_m^{(n)},\;\;
     \frac{u_m^{(n+1)}+p(s_m^{(n+1)})-u_m^{(n)}-p(s_m^{(n)})}{\Delta t}=\frac{u_e(s_m^{(n)})-u_m^{(n+1)}}{\tau},
     \end{eqnarray}
where $s_m^{(n)}=x_{m+1}^{(n)}-x_m^{(n)}$.

The convergence or divergence is actually associated with the stability or instability of the discussed steady-state solution, for which the linear stability condition of (\ref{12}) for the involved equilibrium states are referred to for comparison. However, we should note that not all lengths of these equilibrium states are adequate to apply Eq. (\ref{12}), and that the coupling effect between two adjacent equilibrium states is neglected. Moreover, the perturbation arising from (\ref{30}) to the corresponding steady-state solution can be hardly regarded as \textquotedblleft small". These should give rise to considerable errors in the comparison.

\vspace{0.3cm}
\noindent {\small \hspace{0.0cm} {\bf  Table~1} Comparison between equilibrium values of the steady-state solution and simulated mean values of the semi-discrete model on the divided sections. The initial distribution of (\ref{30}) always converges to the steady-state solution for sufficiently small relaxation time ($\tau=0.03s$); it may diverge from the steady-state solution for large relaxation time ($\tau=0.3s$), which brings about unstable equilibrium states of the solution. }
{{\setlength{\baselineskip}{10pt} \setlength{\parskip}{2pt plus1pt
  minus2pt}} {\small
\begin{center}
\begin{tabular}{|l |c |c |c |c |c |c |c |c |c |c |c |c |}\hline
$$&\multicolumn{4}{|c|}{analytical equilibrium values }&\multicolumn{4}{|c|}{simulated with $\tau=0.03s$}
&\multicolumn{4}{|c|}{simulated with $\tau=0.3s$}\\ \cline{2-13}

$N$&$L_1$&$U$&$L_2$&$D$&$L_1$&$U$&$L_2$&$D$
&$L_1$&$U$&$L_2$&$D$\\ \hline
$250$ &$.1610$&$.2228$&$.1610$&$.1557$ &$.1633$&$.2000$&$.1633$&$.1600$ &$.1633$&$.2000$&$.1633$&$.1600$\\ \hline
$330$ &$.1644$&$.2080$&$.1644$&$.1592$ &$.1648$&$.2133$&$.1633$&$.1600$ &$.1670$&$.2000$&$.1633$&$.1600$\\
 &$.3329$& & &  &$.3326$& & &  &$.3348$& & & \\ \hline
$420$ &$.3329$&$.2080$&$.1644$&$.2297$ &$.3322$&$.2067$&$.1667$&$.2333$ &$.3333$&$.2000$&$.1778$&$.2133$\\
 & & &$.3329$&  & & &$.3333$&  & & &$.3333$& \\ \hline
$550$ &$.3906$&$.2749$&$.3906$&$.2667$ &$.3911$&$.2733$&$.3900$&$.2667$ &$.4111$&$.2667$&$.3433$&$.2467$\\ \hline
$620$ &$.4418$&$.3061$&$.4418$&$.2930$ &$.4400$&$.3067$&$.4433$&$.3000$ &$.5089$&$.2133$&$.3267$&$.2133$\\ \hline
$675$ &$.4824$&$.3285$&$.4824$&$.3124$ &$.4833$&$.3267$&$.4767$&$.3200$ &$.4833$&$.3267$&$.4800$&$.3133$\\
\hline
\end{tabular}\end{center}} }
\vspace{0.3cm}

We choose $\Delta M=1$ to implement a realistic car-following simulation, in which case the particle $m$ represents a vehicle and $s_m(t)$ is headway between the vehicles $m$ and $m+1$. Even $\Delta M$ is not so small, we show the convergence for $\tau$ that is sufficiently small to ensure the stability condition of (\ref{12}) for all $s_0$. Table 1 shows sets of simulated mean density values for comparison with the equilibrium density values of the corresponding steady-state solution that is discussed in Section 3.1. The simulated results are also shown in Fig. 4, by which we clearly observe the convergence. In this case, the semi-discrete model together with its full discrete form of (\ref{31}) works similarly to a relaxation scheme for solving the LWR model of (\ref{13}) (see \cite{Jin:1998,Jin:2000,Herty:2006} for detailed discussion of the relaxation scheme).

However, it is more physically significant to deal with unstable traffic flow in the ring road by taking a much larger relaxation time, by which Eq. (\ref{12}) suggests a narrower region of $s_0$ for the stability of the involved equilibrium states. We take $\tau=0.3s$, which is more realistic when compared with those that are widely used for the car-following model in the literature. With the same values of $N$, the simulated mean values are also given in Table 1, and the simulation results are also shown in Fig. 4 to observe the convergence or divergence.

\begin{center}
{\footnotesize\begin{tabular}{ccc}
 \epsfig{figure=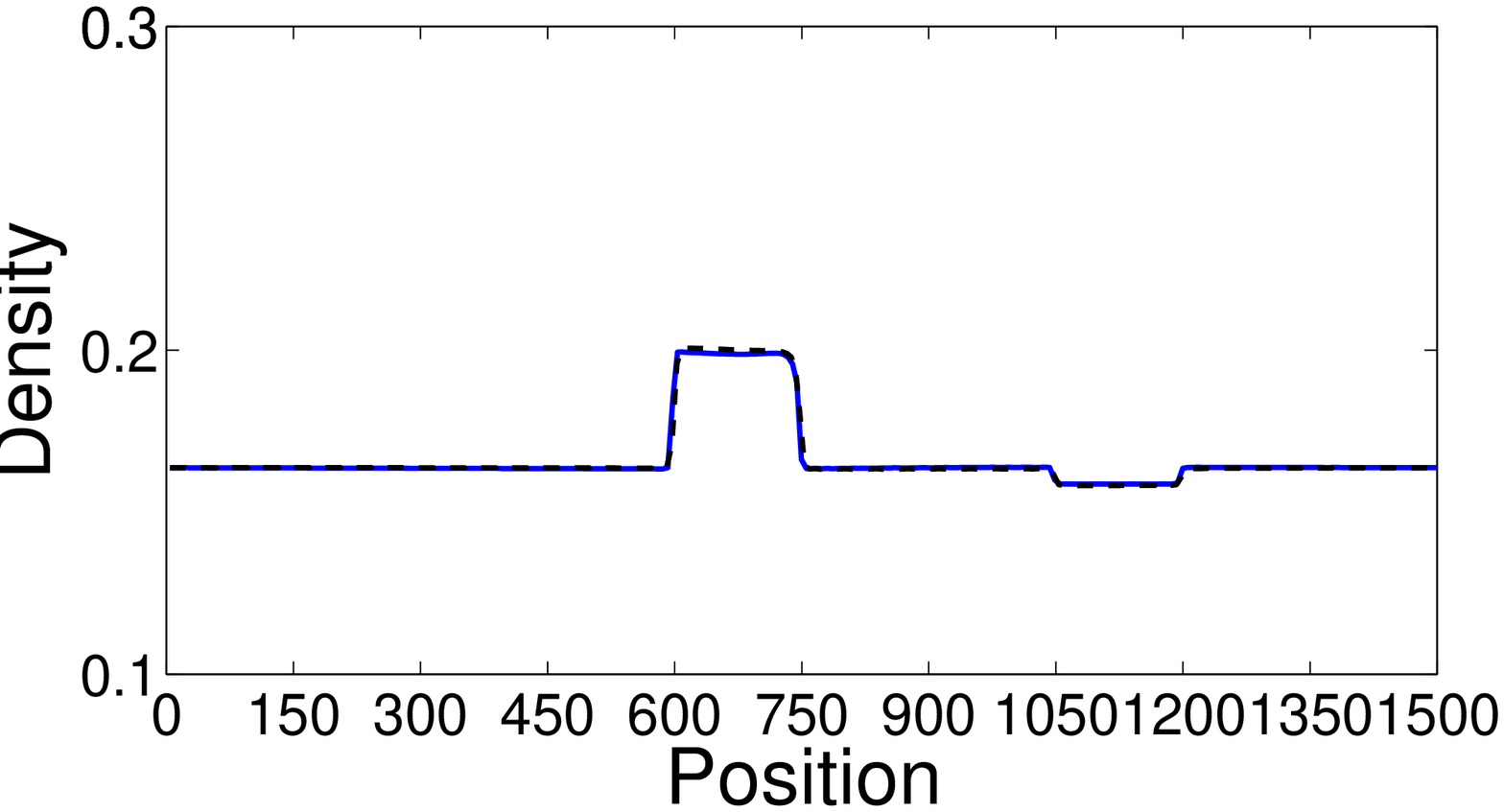,width=2.1 in}  &
 \epsfig{figure=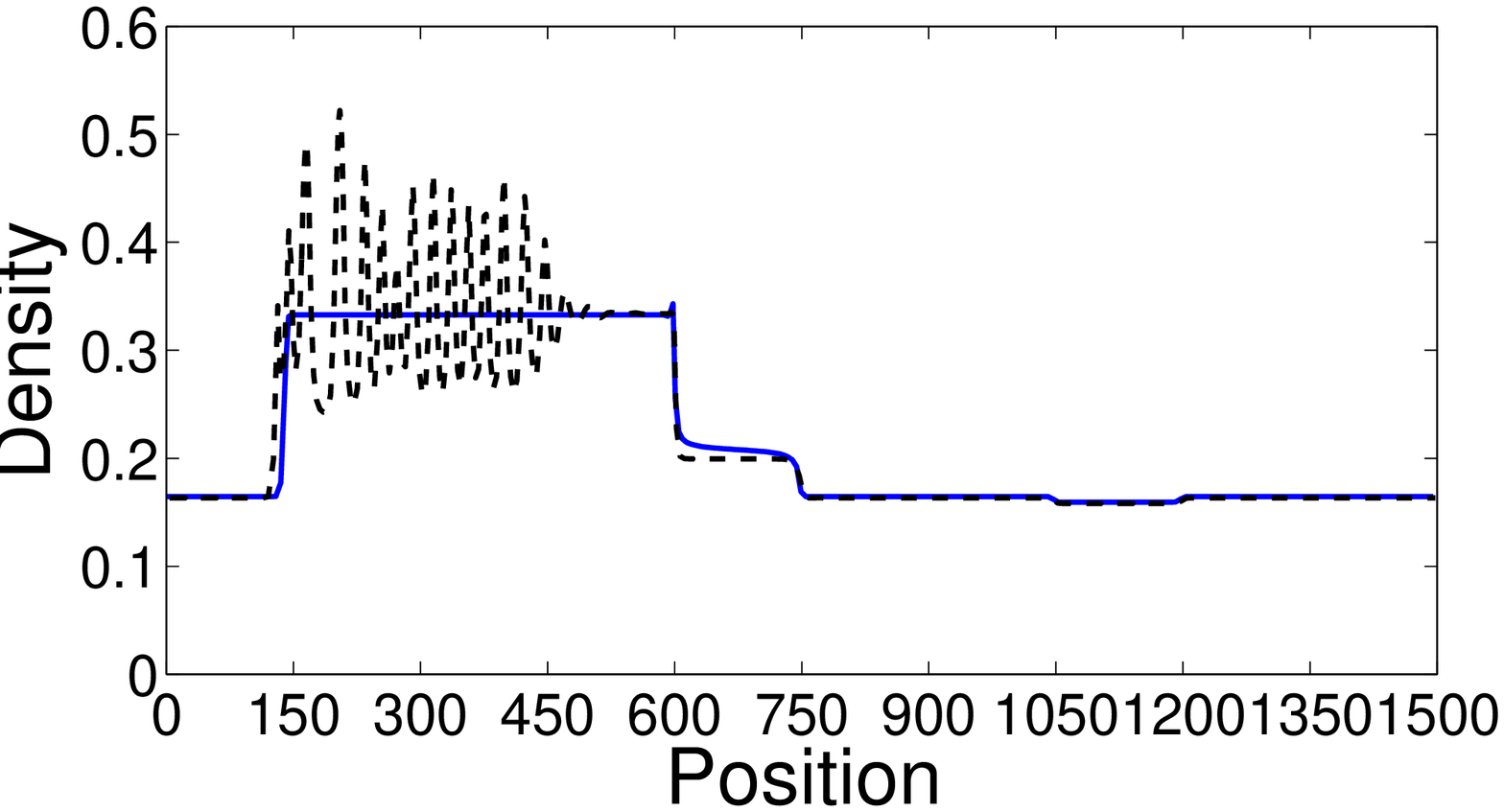,width=2.1 in}  &
 \epsfig{figure=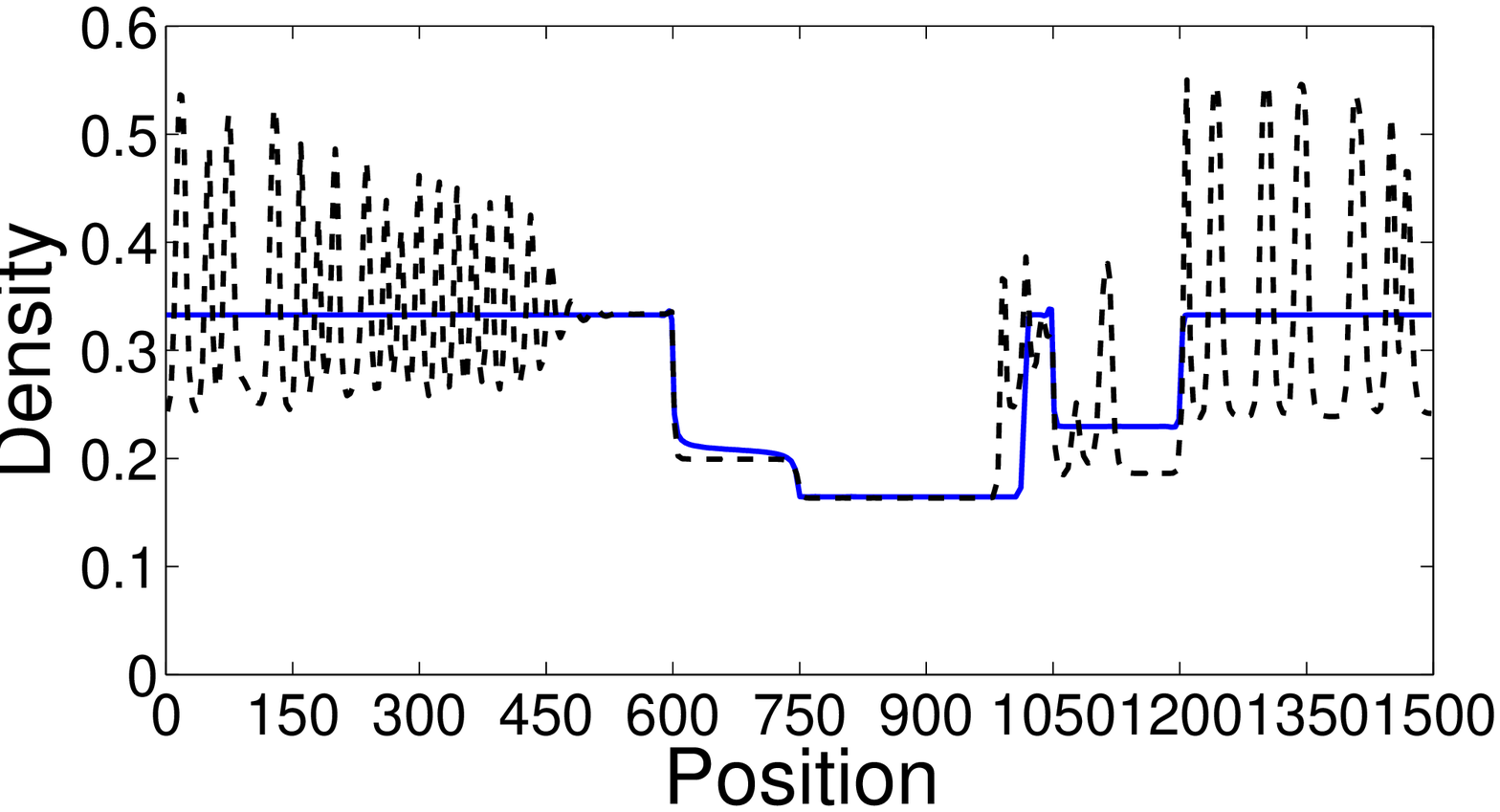,width=2.1 in}  \\
 (a) $N=250$  & (b) $N=330$ & (c) $N=420$ \\
\epsfig{figure=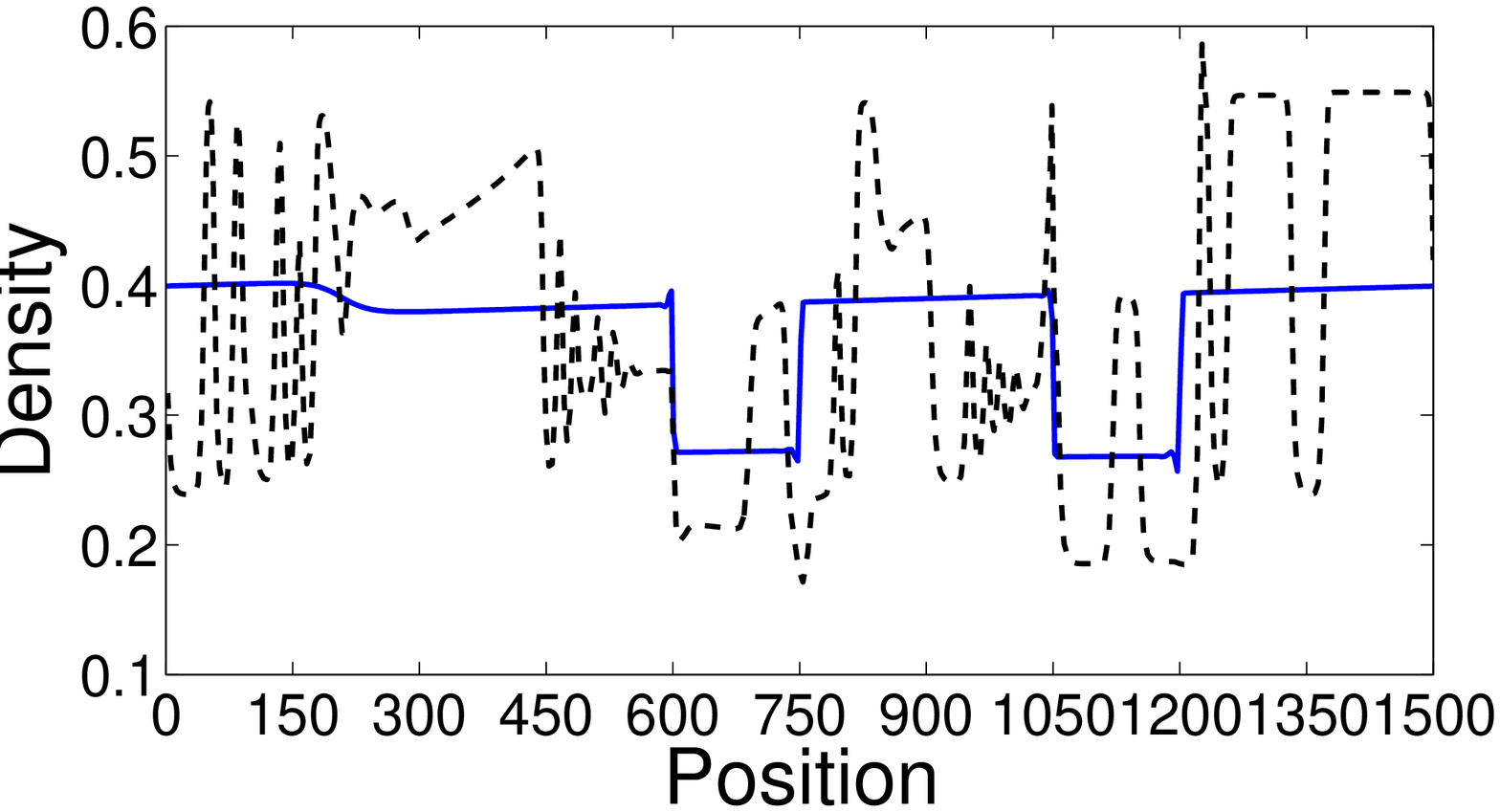,width=2.1 in}  &
 \epsfig{figure=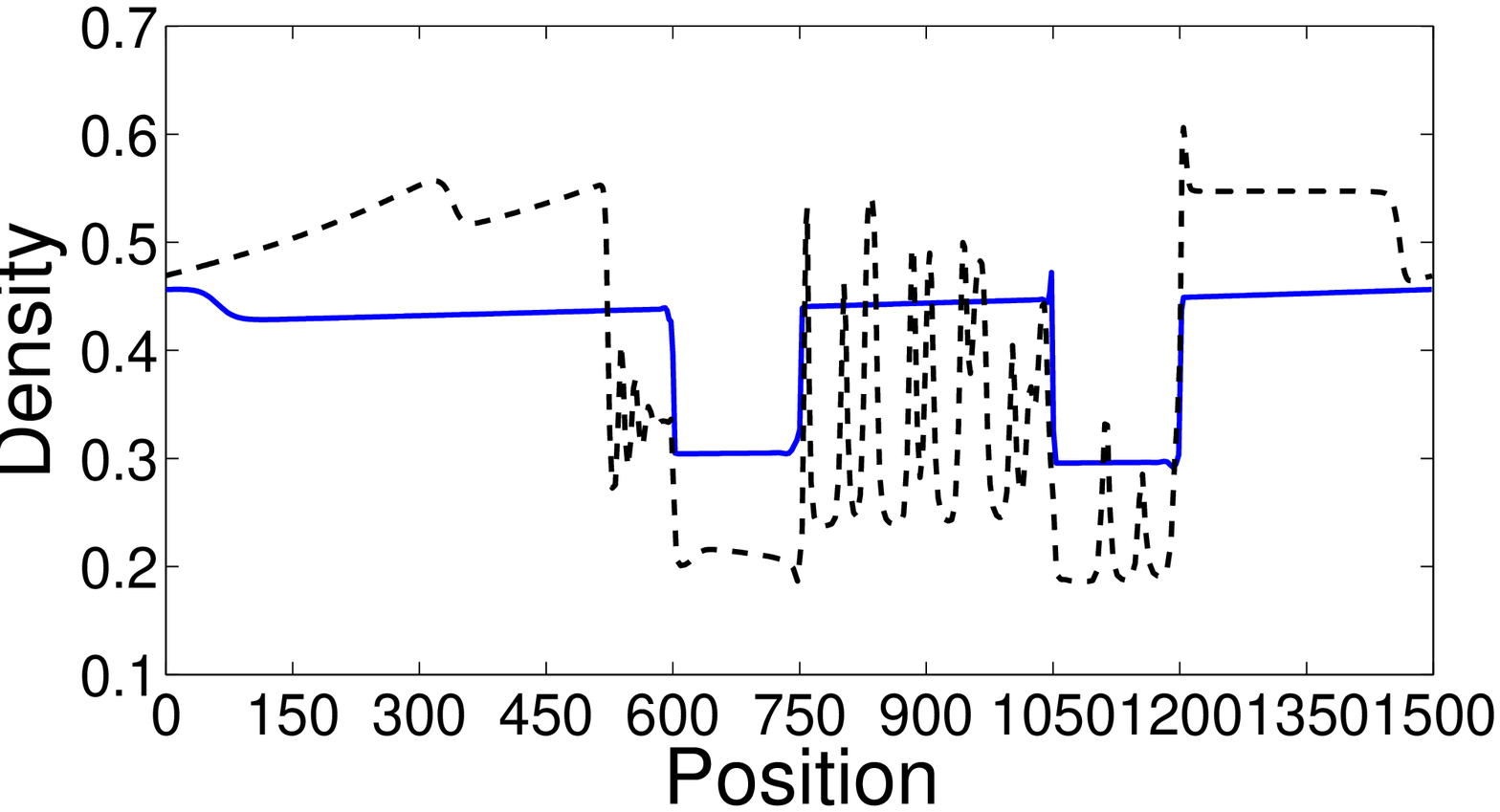,width=2.1 in}  &
\epsfig{figure=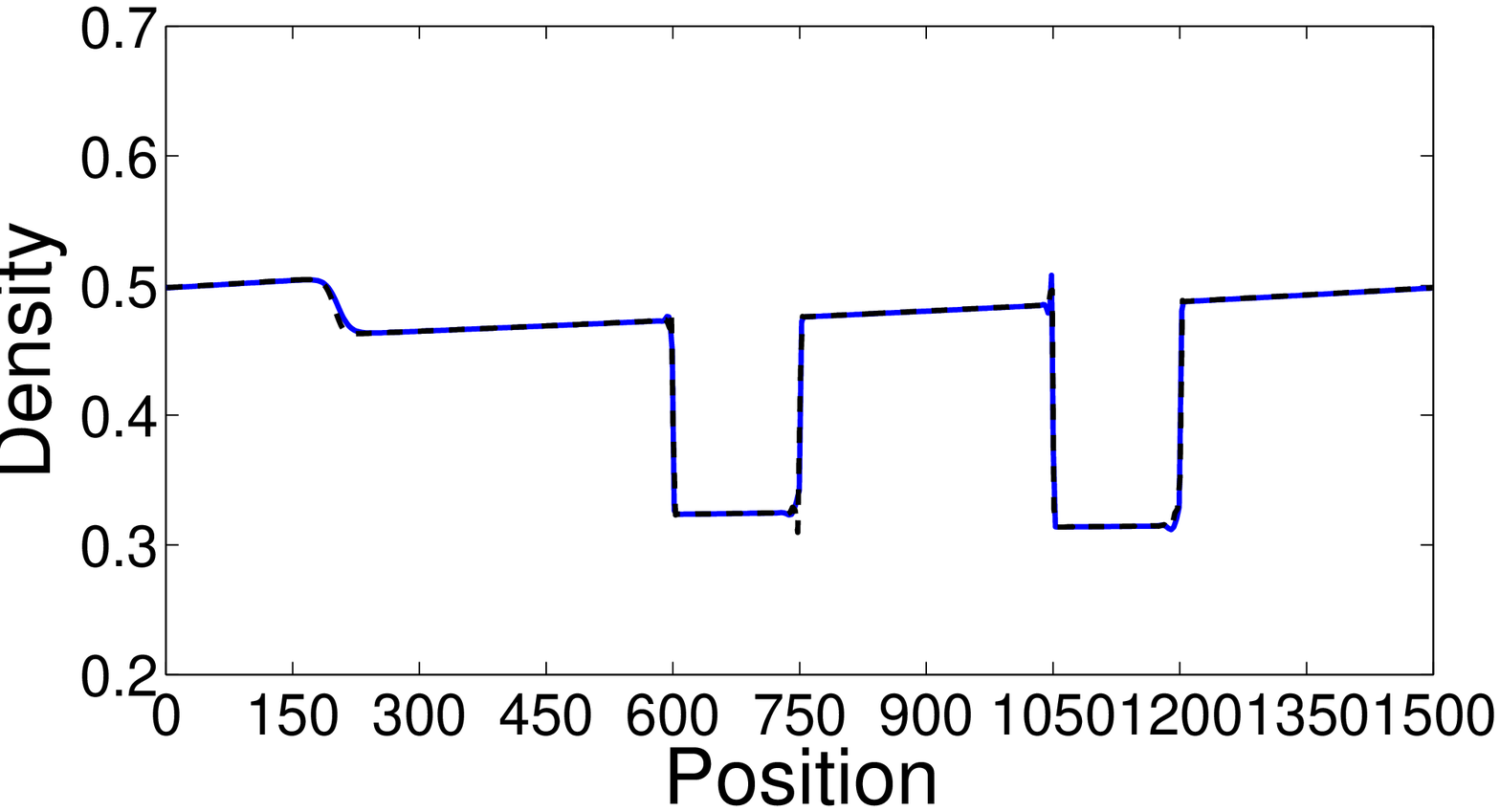,width=2.1 in}  \\
(d) $N=550$ & (e) $N=620$ & (f) $N=675$
\end{tabular}}\\
\vspace{0.3cm}
{\footnotesize  Fig. 4 Evolution of traffic flow on the ring road with initial distribution of (\ref{30}) and for a long simulation time $T$. The density distributions at $T=1500s$, for $\tau=0.03s$ (blue and real line) and $\tau=0.3s$ (black and dotted line). The coordinates of the joints $P_{L_1U}$, $P_{UL_2}$, $P_{L_2D}$, and $P_{DL_1}$ are $600$, $750$, $1050$, and $1200$, respectively}
\end{center}

All involved equilibrium states as well as their stability (for $\tau=0.3s$) depend on the total number $N$ of vehicles in the ring road, which is again taken as a characteristic parameter, and of which the threshold values for changes in stability are combined with those in Eqs. (\ref{24}) and (\ref{27})-(\ref{29}) to further classify the solution (see also Fig. 3). Table 2 indicates the solution properties especially the stability (S) or instability (I) of the equilibrium states with respect to the semi-discrete model, which is according to Eq. (\ref{12}).

Table 2 agrees with Fig. 4 except for some inconsistency in the stability of the equilibrium states of the semi-discrete model, which corresponds to Fig. 4(e). However, both Table 2 and Fig. 4 indicate a similar tendency. That is, the whole solution is quite stable and convergent to the steady-state solution for small or large $N$; the occurrence of instability becomes frequent and intensive for $N$ that is between. This tendency is similar to (but is more complex than) that for a homogeneous level ring road, which was indicated by both theoretical and experimental studies \cite{Kerner:1994,Zhang:2006,Xu:2007,HMZhang:2003,Sugiyama:2008}.

\vspace{0.3cm}
\noindent {\small \hspace{0.0cm} {\bf Table~2} Changes in solution properties arising from changes in $N$. The second column shows the monotone change of $Q_s=Q_s(N)$, where the threshold values $N_{t_1}=253$, $N_{t_2}=404$, $N_{t_3}=415$, and $N_{t_4}=466$ are determined by Eqs. (\ref{21}) and (\ref{23})-(\ref{26}) (also compare with Fig. 3). Other columns show the stability ($S$) or instability ($I$) of the involved equilibrium states by using Eq. (\ref{29}) with $\tau=0.3s$.}
{{\setlength{\baselineskip}{10pt} \setlength{\parskip}{2pt plus1pt
  minus2pt}}{\small
\begin{center}
\begin{tabular}{c c c c c c }\hline
~~~~~~~~$N$~~~~~~~~&~~~~~~$Q_s$~~~~~~&~~~~~~$L_1$~~~~~~&~~~~~~$U$~~~~~~
&~~~~~~$L_2$~~~~~~&~~~~~~$D$~~~~
\\ \hline
    ~$[0,253]$             &$\nearrow$       &$S$     &$S$      &$S$        &$S$\\
    ~$(253,404)$            &$=Q_c$          &$(S,I)$  &$S$      &$S$        &$S$\\
    ~$[404,415)$            &$=Q_c$           &$I$     &$S$      &$S$       &$(S,I)$\\
    ~$[415,466)$            &$=Q_c$           &$I$     &$S$   &$(S,I)$       &$I$\\
    ~$[466,478]$          &$\searrow$      &$I$     &$S$      &$I$        &$I$\\
    ~$(478,579)$          &$\searrow$      &$I$     &$I$      &$I$        &$I$\\
    ~$[579,637)$          &$\searrow$      &$S$     &$I$      &$S$        &$I$\\
    ~$[637,657)$          &$\searrow$      &$S$     &$S$      &$S$        &$I$\\
    ~$[657,1500]$         &$\searrow$      &$S$     &$S$      &$S$        &$S$\\
\hline
\end{tabular}\end{center}} }
\vspace{0.3cm}

Most oscillations in Figs. 4(b)-(e) represent stop and go waves in traffic. Among these, the waves in Figs. 4(b)-(c) demonstrate propagations only within one or more road sections, while we observe steady-state traffic flow in other road sections through comparison with the simulated results for $\tau=0.03s$. In contrast, the waves in Figs. (d)-(e) propagate throughout the ring-road, in which case no steady states are locally observed.  On the other hand, we see some regularity of the backward moving stop-and-go waves, which are somewhat similar to (but not as regular as) those traveling waves that were studied under homogeneous ring road conditions \cite{Kerner:1994,Greenberg:2004,Zhang:2006,Xu:2007,Zhang:2012}. The described solution properties change little for much longer simulations. Fig. 5 shows the evolution process for $N=550$, which corresponds to Fig. 4(d).

\begin{center}
\epsfig{figure=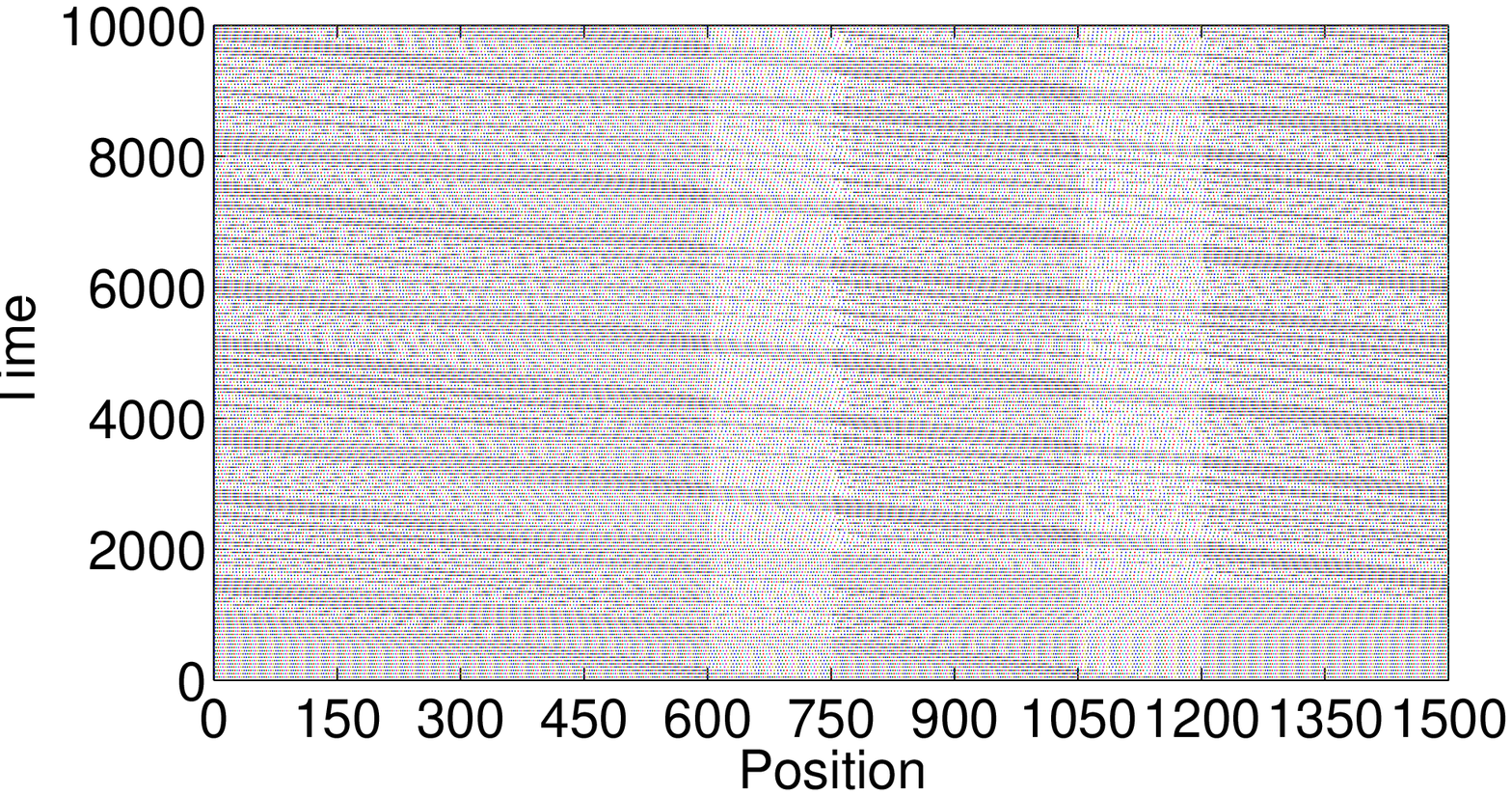, width=3.8in} \\
 {\footnotesize
 Fig. 5 Evolution of the initial distribution of (\ref{30}) with $N=550$ and for $t\leq 1500s$, which indicates stop-and-go waves moving backward throughout the ring-road.
 }
\end{center}

\section{Conclusions}
We thoroughly investigate the steady-steady solution on an inhomogeneous ring road using a semi-discrete model.  According to the analysis of the wave pattern at a dividing point between two equilibria, the steady-state solution is uniquely determined. Moreover, the stability of the solution is related to that of the involved equilibria.

We find that both the wave types and the stability of the steady-state solution depend on the total number $N$ of vehicles on the ring road, and that the simulation results agree with the analysis in general. For the case with a realistic relaxation time $\tau=0.3s$, a casually distributed traffic state (of Eq. (\ref{30})) is able to stably evolve into the corresponding steady-state flow for light or heavy traffic with small or large number $N$, but liable to generate stop-and-go waves for congested traffic with the number $N$ that is between. This tendency is similar to that on a homogeneous level ring road, and the indicated phenomena are physically significant for a better understanding and the management of traffic flow on an inhomogeneous road in general.

However, we still need more efficient tools for stability analysis of the discussed steady-state solution due to the inadequate lengths of and the neglect of interaction between the involved equilibrium
states in the current analysis.


\bigskip

\noindent \Large \textbf{Acknowledgements} \normalsize

The study was jointly supported by grants from the National Natural Science Foundation of China (11072141,11272199), the National Basic Research Program of China (2012CB725404), Shanghai Program for Innovative Research Team in Universities, the Research Grants Council of the Hong Kong Special Administrative
Region, China (Grant No. HKU7184/10E), and a National Research Foundation of Korea grant funded by the Korean government (MEST) (NRF-2010-0029446).

\bigskip
{\small \setlength{\baselineskip}{10pt} \setlength{\parskip}{2pt
plus1pt
  minus2pt}

\end{document}